\documentclass[11pt,preprint]{aastex}
\shorttitle{Ultra Violet spectral indices}
\shortauthors{Serven et al.}
\usepackage{graphicx}
\usepackage{epstopdf}
\usepackage{float}
\begin{document}
\title{NH and Mg Index Trends in Elliptical Galaxies}
\author{Jedidiah Serven$^1$, Guy Worthey$^1$, Elisa Toloba$^2$ \& Patricia S{\'a}nchez-Bl{\'a}zquez$^3$}
\affil{$^1$Department of Physics and Astronomy, Washington State University,Pullman, WA 99164-2814}
\affil{$^2$Universidad Complutense de Madrid, 28040, Spain}
\affil{$^3$Instituto de Astrof'sica de Canarias, La Laguna, E-38200 Tenerife, Spain}
\begin{abstract}

We examine the spectrum in the vicinity of the NH3360 index of \cite{Davidage94}, which was  defined to measure the NH absorption around 3360 \AA\/ and which shows almost no trend with velocity dispersion \citep{Toloba09}, unlike other N-sensitive indices, which show a strong trend \citep{Graves07}. Computing the effect of individual elements on the integrated spectrum with synthetic stellar population integrated spectra, we find that, while being well correlated with nitrogen abundance, NH3360 is almost equally well anti-correlated with Mg abundance. This prompts the definition of two new indices, Mg3334, which is mostly sensitive to magnesium, and NH3375, which is mostly sensitive to nitrogen. Rather surprisingly, we find that the new NH3375 index shows a trend versus optical absorption feature indices that is as shallow as the NH3360 index. We hypothesize that the lack of a strong index trend in these near-UV indices is due to the presence of an old metal-poor component of the galactic population. Comparison of observed index trends and those predicted by models shows that a modest fraction of an old, metal-poor stellar population could easily account for the observed flat trend in these near-UV indices, while still allowing substantial N abundance increase in the larger galaxies.

\end{abstract}
\section{Introduction}
One of the biggest problems with any attempt to determine the chemical make up of a stellar system is trying to disentangle the effects of C, N, and O in any given spectrum. This is due to the fact that in cool stars and elliptical galaxies, C, N, and O show themselves in any spectrum almost exclusively through molecular species such as NH, CN, C$_2$, CH, and CO rather than atomic species. The intertwining of these three elements starts with CO. The fact that CO has the highest dissociation energy of these molecules means that CO will form the most prolifically if given enough O and C. Since O is typically the most abundant of these three elements, C becomes incorporated into CO instead of the other molecular species. The rest of these molecules are connected through balancing of molecular equilibria. These interactions give a net effect of O acting like anti-CN since adding any O will decrease the amount of C available for the formation of other molecules \citep{serv05}.

To begin disentangling these three elements it should be possible to start with pseudo equivalent width indices that are sensitive to these elements such as C$_2$4668 \citep{wor94b}, which is sensitive to C, and the indices CO5161 and CO4685 \citep{serv05} which are insensitive to N but react to C and O. After getting a grip on the C and O abundances, one can use the CN band to determine the N abundance. Unfortunately, disentangling C, N, and O this way turns out to be difficult \citep{bur03}. To low precision, most previous results agree that Mg, C, N, and Na appear to be enhanced in large elliptical galaxies and also correlated with velocity dispersion \citep{wor98a,sanchez03,Kelson06,Graves07}.

What is most often suggested as an alternative for determining the N abundance is the use of the NH feature at 3360 \AA\ since, as has been noted before, it is insensitive to C and O \citep{sne73,nor02} and it is also directly and sensitively measuring N abundances \citep{bes82,tom84}.

Below, we will show that this is not the whole story and that there are other contributors to the NH feature. Unfortunately, until recently there had been fewer than 15 early-type galaxies with published NH3360 values due in large part to relative insensitivity of detectors in the near-UV \citep{Toloba09}. So making use of the NH3360 feature was almost impossible until the introduction of NH3360 values of 35 early-type galaxies from \cite{Toloba09}. 

In \cite{Toloba09} this sample of 35 galaxies was measured using indices NH3360 \citep{Davidage94}, CNO3862, CNO4175, CO4685 \citep{serv05}, and Mg $b$ \citep{bur84}. Their findings included that there exists a flat relation between the NH3360 index and velocity dispersion. This seems to indicate that there does $not$ exist a velocity dispersion relation for nitrogen, contrary to work done previously. For example, since the CN relation is stronger and tighter that the C$_2$4668 relation, it seems that there should be a positive [N/Fe] trend with velocity dispersion \citep{trag}.

It is toward attempting to explain this disparity that the rest of this paper is aimed. In the first section, the response of the NH3360 feature to various elements is calculated as in \cite{serv05} using a greatly improved version of those models. Then, after seeing Mg contamination in the index, two new indices are defined, one for the contaminate Mg, and the other for N. The responses for these new indices are then calculated. In \S 3 the NH3360 index along with NH3375, Mg3334, CN$_1$ and Mg $b$ are plotted and compared to models to determine if the effects of an old metal-poor stellar population can explain the observed index trends. Lastly, the results and conclusions are discussed in \S 4.

\section{Analysis}
The method for determining the response of NH3360 as well as the two new indices was similar to that used in \cite{serv05}. In \cite{serv05} simple models of a galaxy spectrum were constructed, using a G dwarf and a K giant. These models varied in that there was one base model of solar metallicity and then 23 variations, each one with a particular elemental abundance doubled. Then, the ratios of these spectra were taken to find any spectral influence due to any particular element.

The only difference is that the numbers tabulated here come from measuring these indices in full single stellar population models. These models are a version of the (\cite{wor94}; \cite{trag}) models that use a grid of synthetic spectra in the optical \citep{lee}. Age, overall metallicity Z, and 23 individual elemental abundances can be varied independently in the models, and spectra for single-burst (simple) stellar populations produced. We plot results based on \cite{wor94} isochrones, in this paper, but confirm all results with Padova \citep{bert94} evolution. The grid of synthetic spectra is complete enough to predict nearly arbitrary composition.

\begin{figure}[H]
\includegraphics[width=5in]{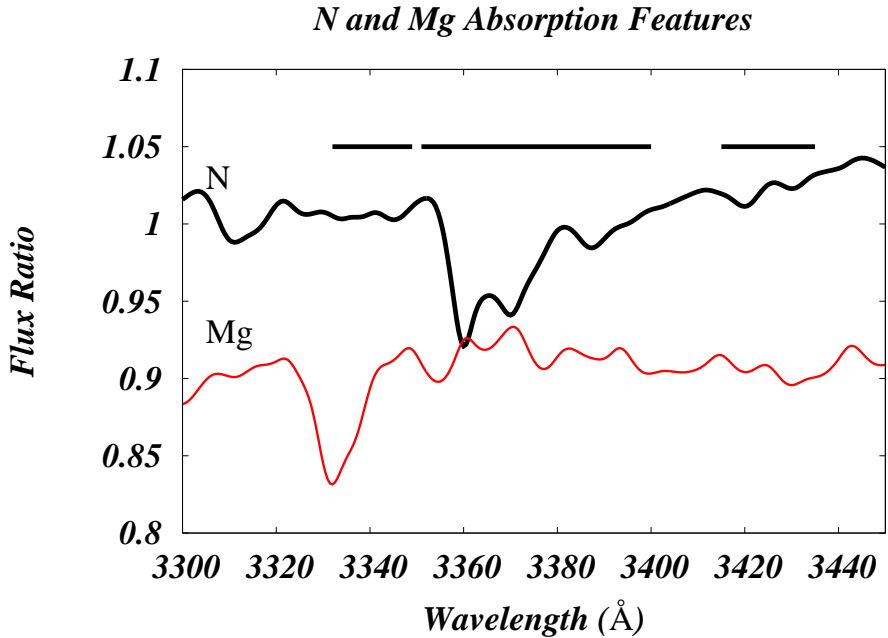}
\caption{NH3360 index (horizontal lines). The relative spectral responses due to 0.3 dex increase in N (black) and Mg (red) in the same spectral region are shown. The feature of interest lies in the blue continuum pass band for NH3360 which overlaps with Mg3334 index feature. Note that a gap between the blue passband and continuum passband has been added for clarity.
\label{Fig1} }
\end{figure}

The responses of NH3360, NH3375, Mg3334, Fe4383, Mg $b$ and H$\beta$ can be found in Table 7. What can be seen is that the NH3360 index, while indeed insensitive to C and O, is sensitive to Mg and to a lesser degree Fe and Ni. The anti-correlation with Mg is due to a small Mg absorption feature located in the blue continuum passband (Fig. \ref{Fig1}). In order to remove this Mg dependence, a new index was defined; index NH3375, with an altered blue pseudocontinuum definition that avoids this feature. The new index NH3375 response, which is also found in Table 7, shows a smaller dependence on Mg with some small dependence on Ti and Ni. Unfortunately, the sensitivity of NH3375 to Ti and Ni comes from features of these two elements that overlap the NH feature itself. This limits the degree to which these sensitivities can be removed from any index definition.

\begin{figure}[H]
\includegraphics[width=5in]{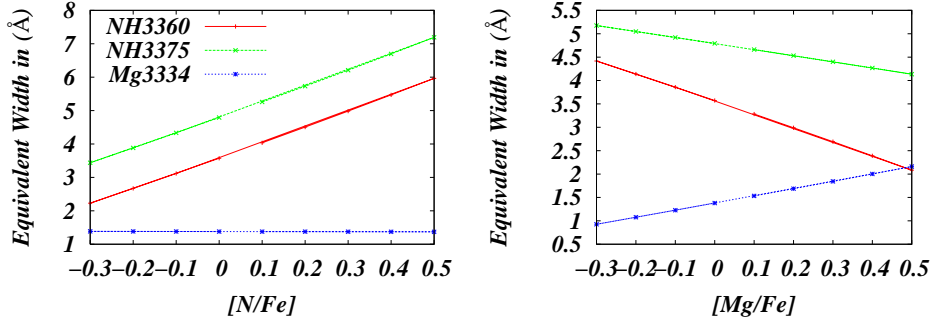}
\caption{In the first panel NH3360 (red), NH3375 (green), Mg3334 (blue) against [N/Fe]. In the second panel NH3360 (red), NH3375 (green), and Mg3334 (blue) against [Mg/Fe]. This figure illustrates the sensitivity of these indices to changes in the N and Mg abundance at fixed [Z/H] = [Fe/H] = 0.0 and age = 8 Gyr with [Mg/Fe] = 0.0 for the first panel and [N/Fe] = 0.0 for the second.
\label{Fig2} }
\end{figure}

For the sake of complementarity, a new index was defined for the small Mg feature as well; index Mg3334. The response of this index to Mg abundance is surprisingly clean, showing only small sensitivity to O (See Table 6). This is most likely due to the effects of molecular equilibrium involving molecular species formed for C, N and O. The definitions of these new indices, as well as NH3360, can be found in Table \ref{Table1}. The index sensitivities to N and Mg are modeled in Figure \ref{Fig2}.

\begin{table}[H]
\begin{center}
\begin{tabular}{ l c c c c}
\multicolumn{5}{c}{Table 1} \\
\hline
\hline
Index & Blue Passband & Index Passband & Red Passband & Reference\\
\hline
NH3360 & 3332-3350 & 3350-3400 & 3415-3435 & \cite{Davidage94}\\
NH3375 & 3342-3352 & 3350-3400 & 3415-3435 &  This work \\
Mg3334 & 3310-3320 & 3328-3340 & 3342-3355 &   This work \\
Mg $b$  & 5142.625-5161.375 & 5160.125-5192.625 & 5191.375-5206.375 & \cite{trag}\\
\hline
\end{tabular}
\caption{Index passband definitions in \AA\/.}
\label{Table1}
\end{center}
\end{table}

A Z versus age sensitivity parameter (Zsp) was calculated for NH3360, NH3375, Mg3334, and Mg $b$ as it was in \cite{wor94}. The Zsp is the ratio of the percentage change in age to the percentage change in Z ( Z $\approx 0.01689 \times 10^{[Fe/H]}$) of the index measured as shown below. 

\begin{equation}
 Zsp = {[\delta \textrm{I}_m/\delta \textrm{log(Z)}]\over[\delta \textrm{I}_a/\delta \textrm{log(age)}]}
\end{equation}
 
Here $\delta$$\textrm{I}_m$/$\delta$log(Z) is the partial derivative of the index with respect to metallicity at age = 12 Gyrs. Similarly, $\delta$$\textrm{I}_a$/$\delta$log(age) is the partial derivative of the index with respect to age at solar metallicity. These sensitivities are shown in Table \ref{table2}. Note that the models indicate that both NH3360 and NH3375 are far more sensitive to age than they are to metallicity, especially NH3360 which has almost no metallicity sensitivity in the metal rich regime, while the Mg indices are more sensitive to metallicity.

\begin{table}[H]
\begin{center}
\begin{tabular}{ l c }
\multicolumn{2}{c}{Table 2} \\
\hline
\hline
Index & Zsp \\
\hline
NH3360  &  0.2 \\
NH3375  &  0.6 \\
Mg3334  &  1.1 \\
Mg $b$  &  1.7 \\
\hline
\end{tabular}
\end{center}
\caption{This shows the Z versus age sensitivity parameter (Zsp), which gauges how changes in metallicity and age effect various indices. A large Zsp indicates a larger dependance on the overall metallicity than on age with 1.0 indicating that age and metallicity effect the index equally.}
\label{table2}
\end{table}

\begin{figure}[H]
\includegraphics[width=5in]{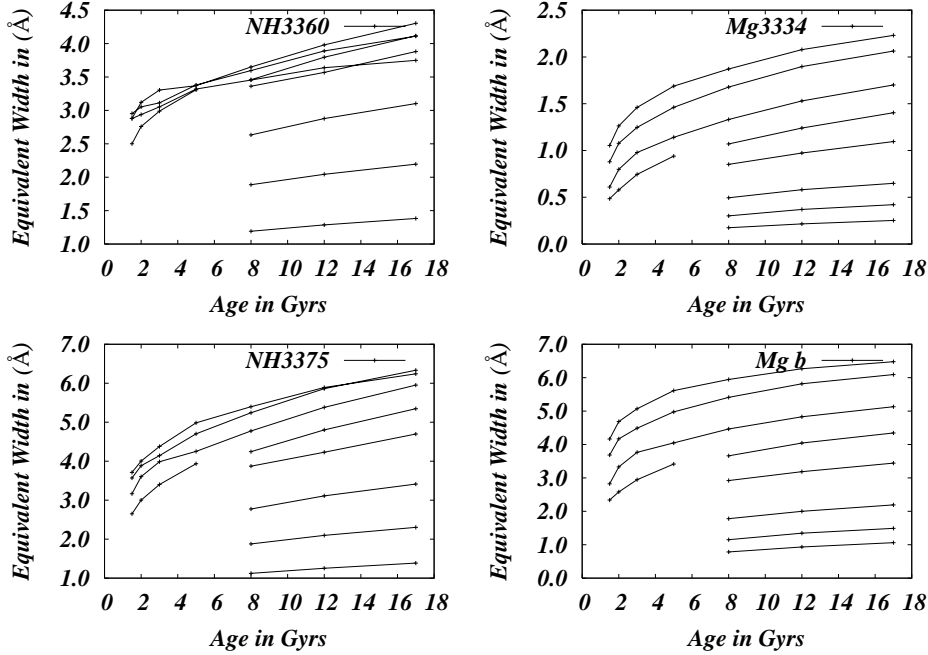}
\caption{NH3360, NH3375, Mg3334, and Mg $b$ against age for various metallicities. The NH3360 index (top left), the NH3375 index (bottom left), the Mg3334 index (top right), and the Mg $b$ index (bottom right) are plotted for metallicities from top to bottom $0.5, 0.25, 0, -0.225($ages $ 1.5, 2, 3, 5),-0.25($ages $8, 12, 17), -0.5, -1, -1.5, -2$ and ages 1.5, 2, 3, 5, 8, 12 and 17 Gyrs.
\label{Fig3} }
\end{figure}

A plot illustrating the sensitivity of these indices to metallicity and age can be found in Figure \ref{Fig3}. Note that the NH3360 index, although age sensitive, behaves much the same for metallicities larger than  Z $=-1$. The other indices exhibit behavior of increasing and plateauing out over time.

\section{Observations and Results}

The \cite{Toloba09} sample consists of long-slit spectra for 35 elliptical galaxies collected with the 4.2 m William Herschel Telescope at Roque de los Muchachos Observatory, using the ISIS spectrograph. The spectra were extracted within a central equivalent aperture of 4$^{\prime\prime}$. The wavelength coverage is from 3140 to 4040 \AA\ with a resolution of 2.3 \AA\ (FWHM) and a typical signal to noise of $S/N=40$ per \AA\/. These spectra were chosen as a subset of those presented in \cite{sanchez06} so that this near-UV data could be supplemented with optical data in the wavelength range 3500 to 5250 \AA\/. The galaxies in this set were also chosen to include field, Virgo and Coma cluster ellipticals, which cover a range of velocity dispersions (130 $<  \sigma < $ 330 km s$^{-1}$).

From this sample the NH3360, NH3375, Mg3334, CN$_1$, Mg $b$ and H$\beta$ indices were measured at a velocity dispersion of 200 km s$^{-1}$ and corrected for the effects of velocity dispersion of the galaxy using synthetic spectra if the galaxies were bigger than that. The measurements for NH3360, NH3375 and Mg3334 were taken from the \cite{Toloba09} spectra, while the measurements for CN$_1$, Mg $b$ and H$\beta$ were taken from the \cite{sanchez06} optical spectra. These measurements were then plotted against the Fe4383 index to compare the trends of these near-UV indices which are sensitive to N and Mg with the slightly redder CN$_1$ and Mg $b$ which are also sensitive to N and Mg (Fig. \ref{fig4}). For comparison, Mg3334 vs Mg $b$ and H$\beta$ vs Fe4383 are also plotted. The trend lines shown in Figure 4 are best-fit lines (Table 4) calculated using fitexy.f \citep{numrec}, a program for finding the best-fit line for data with errors in both the x and y coordinates. It minimizes the distance of each point form the line while taking into account weighting by the precision of the individual measurements in both the x and y coordinate.

The index plotted in the first panel of Fig. \ref{fig4} is NH3360, in the second NH3375 and in the third is Mg3334, all against the Fe4383 index. The index plotted in the fourth panel of Fig. \ref{fig4} is CN$_1$, in the fifth is Mg $b$ both against the Fe4383 index, in the sixth is a plot of Mg3334 vs Mg $b$ and in the seventh is a plot of H$\beta$ vs Fe4383. In each panel along with the index are plotted three lines. These lines represent the respective indices measured from composite 12 Gyr galactic model of metallicity [Fe/H] = 0.0 and [Fe/H] = 0.25. The red line is a model of solar metallicity with no subpopulation. The green line is a model of solar metallicity, but with 5 percent of the galactic mass consisting of a 12 Gyr, metal-poor ([Fe/H] = $-$1.5) subpopulation. The blue line is the same base model except that the subpopulation comprises 10 percent of the total mass. The metal-poor subpopulation has a blue horizontal branch morphology.

What can be seen in Fig. \ref{fig4} is that, for the two NH indices the index trends tend to decrease and Mg3334 looks flat in contrast to the panels 4, 5 and 7 of Fig. \ref{fig4} which show a definite trend with increasing metallicity. What can also be seen is that, for the galactic models as the percent mass of the subpopulation is increased (red to green to blue) the model index trends flatten out; twisting to come into fairly good agreement with the observed index trends within a metal-poor subpopulation fraction somewhere around 5 percent by mass. The redder indices of CN$_1$, Mg $b$ and H$\beta$ show that the subpopulation has little to no effect on these index trends except lowering the average metallicity.

Further evidence for the existence of an old metal-poor population can be see in panel 6 of Fig. \ref{Fig3}, what can be seen here is that Mg3334 and Mg $b$ which are fairly clean indicators of Mg (\cite{serv05}; This Work) do not appear to be measuring the same abundance trend. This is an indication that the near-UV indices do indeed suffer from line strength dilution due to an underlying bright and weak-lined near-UV population. The data fits slope and model slopes are listed in Table 4, while the slope errors are relatively large it is evident from the numbers that a composite population with a $5-10\%$ metal-poor subpopulation fits the observed trend better.

\begin{figure}[H]
\includegraphics[width=7in]{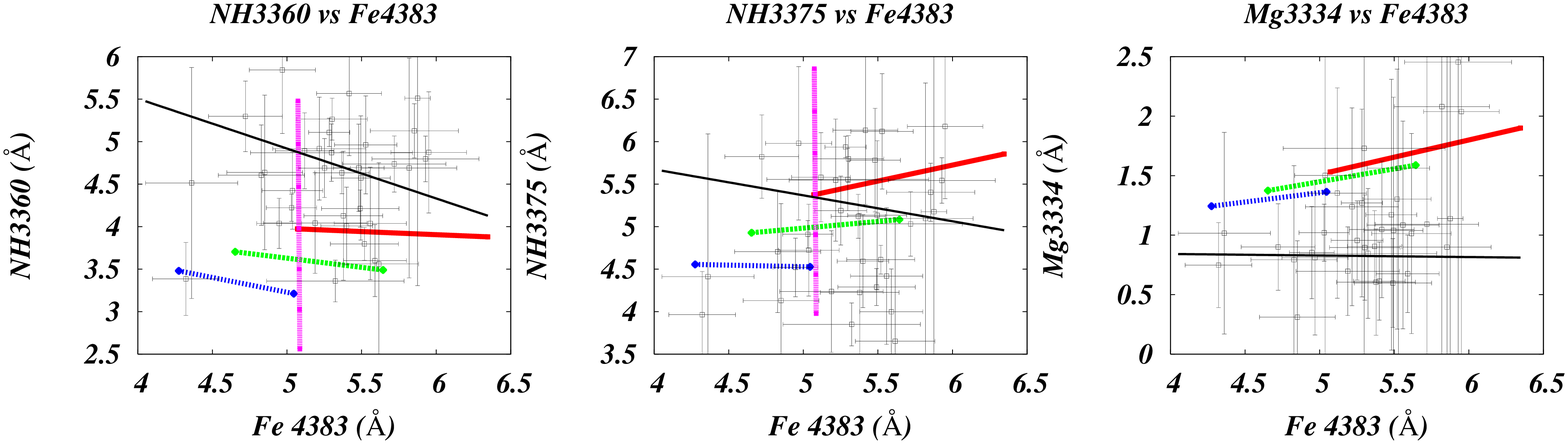}
\includegraphics[width=7in]{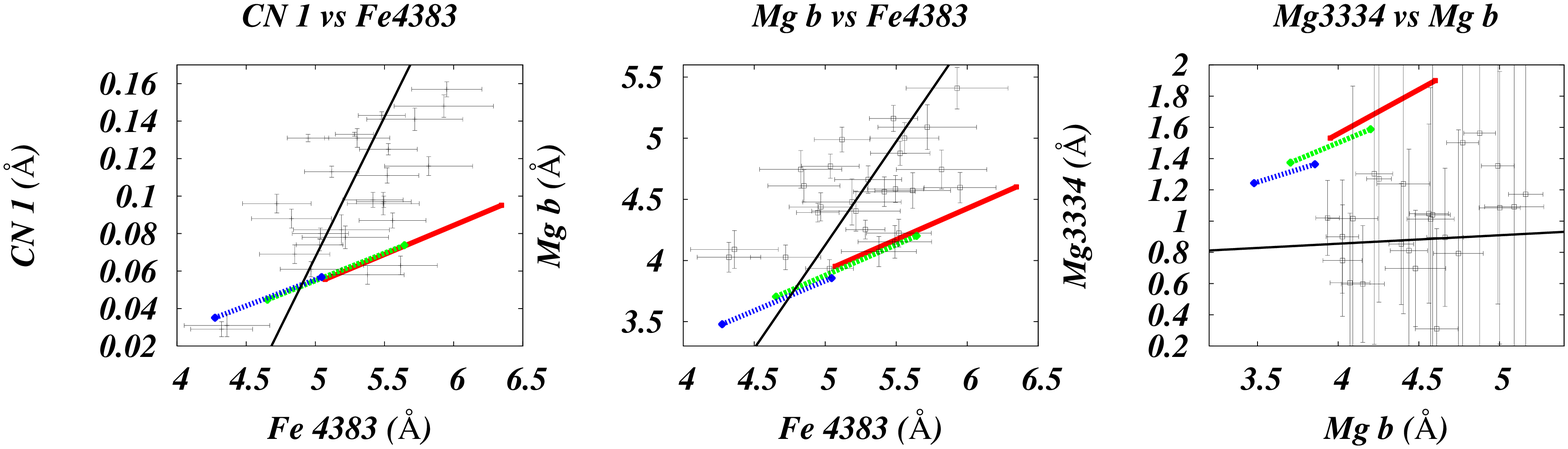}
\includegraphics[width=2.5in]{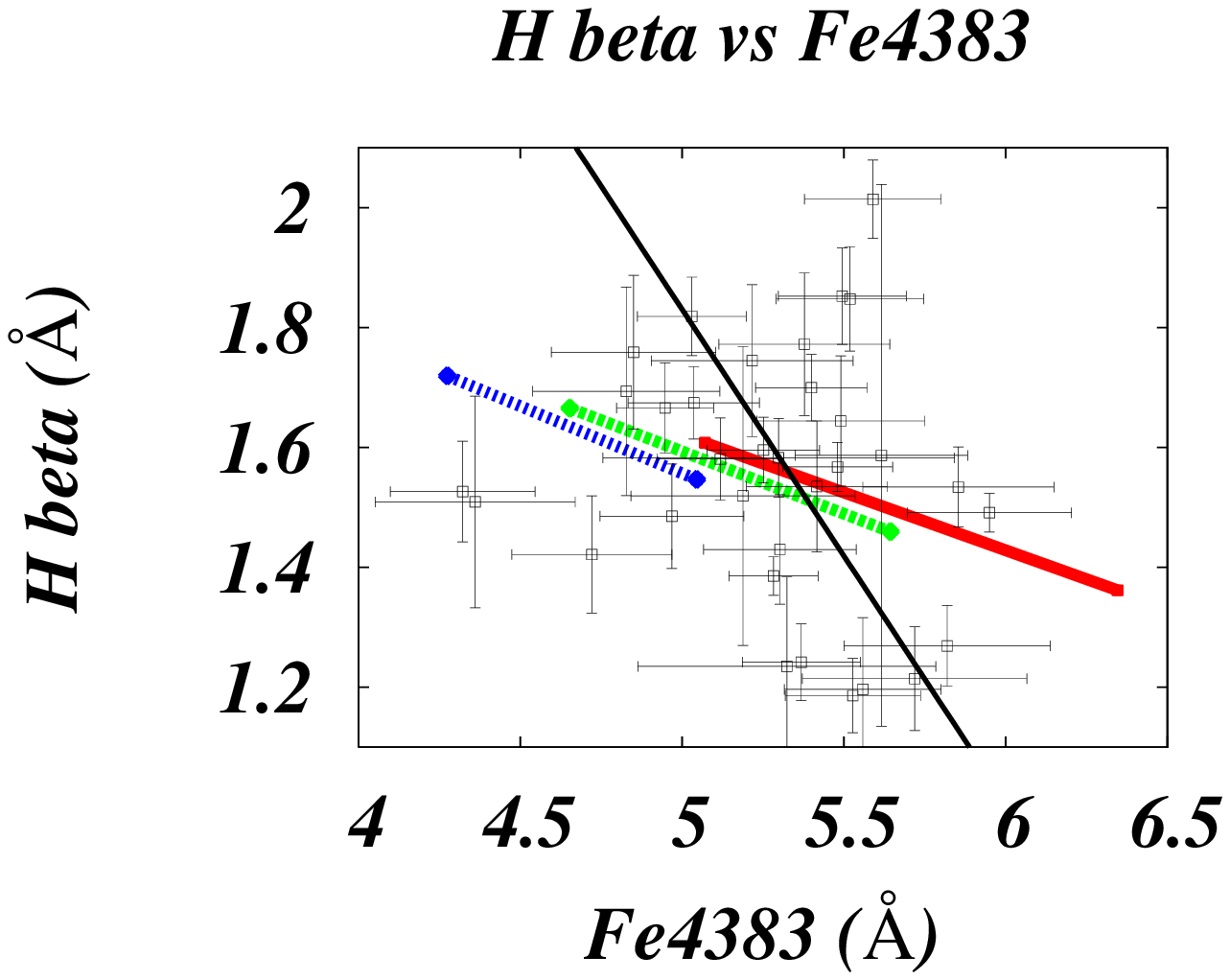}
\caption{In the first, second and third panels, NH3360, NH3375 and Mg3334 indices vs Fe4383 are plotted, respectively. In the fourth and fifth panels, CN$_1$ and Mg $b$ indices vs Fe4383 are plotted respectively. In the sixth panel Mg3334 vs Mg $b$ is plotted and in the seventh panel H$\beta$ vs Fe4383 in plotted. In each panel along with the index are plotted three lines. These lines represent the respective indices measured from a 12 Gyr galaxy model of metallicity [Fe/H] = 0.0 to a metallicity of [Fe/H] = 0.25. The red line is a model of solar metallicity with no subpopulation. The green line is a model of solar metallicity, but with 5 percent of the galaxy mass consisting of a 12 Gyr, metal-poor ([Fe/H] = $-$1.5) subpopulation. The blue line is the same model except that the subpopulation is now 10 percent of the total mass and the black lines are fits to the data. The data sources are \cite{Toloba09} and \cite{sanchez06}.
\label{Fig4} }
\end{figure}

To compare the relative effects of N abundance increase and overall metallicity, the NH3360 vs CN$_1$ index measurements were plotted along with the subpopulation models and a 12 Gyr solar metallicity model of varying N abundance (pink line) (Fig.\ref{fig4}). In Figure 5, it can be seen from comparison of both models and data that neither account for the observed trend alone. However, a combination of increasing N abundance and the presence of a metal-poor subpopulation could reproduce the observed trend if for example the higher metallicity (larger CN$_1$) models also had a N enhancement relative to the lower metallicity model. Then for a modest N enhancement of $\approx$ 0.04, 0.06 and 0.07 dex for the no subpopulation, $5\%$ and $10\%$ models respectively the model trend would concur with the observed slope. This is not a definitive statement about the relative amounts of these quantities, owing to the fact that there may be interplay between the N content and the presence of a metal-poor population. 

\begin{figure}[H]
\includegraphics[width=4in]{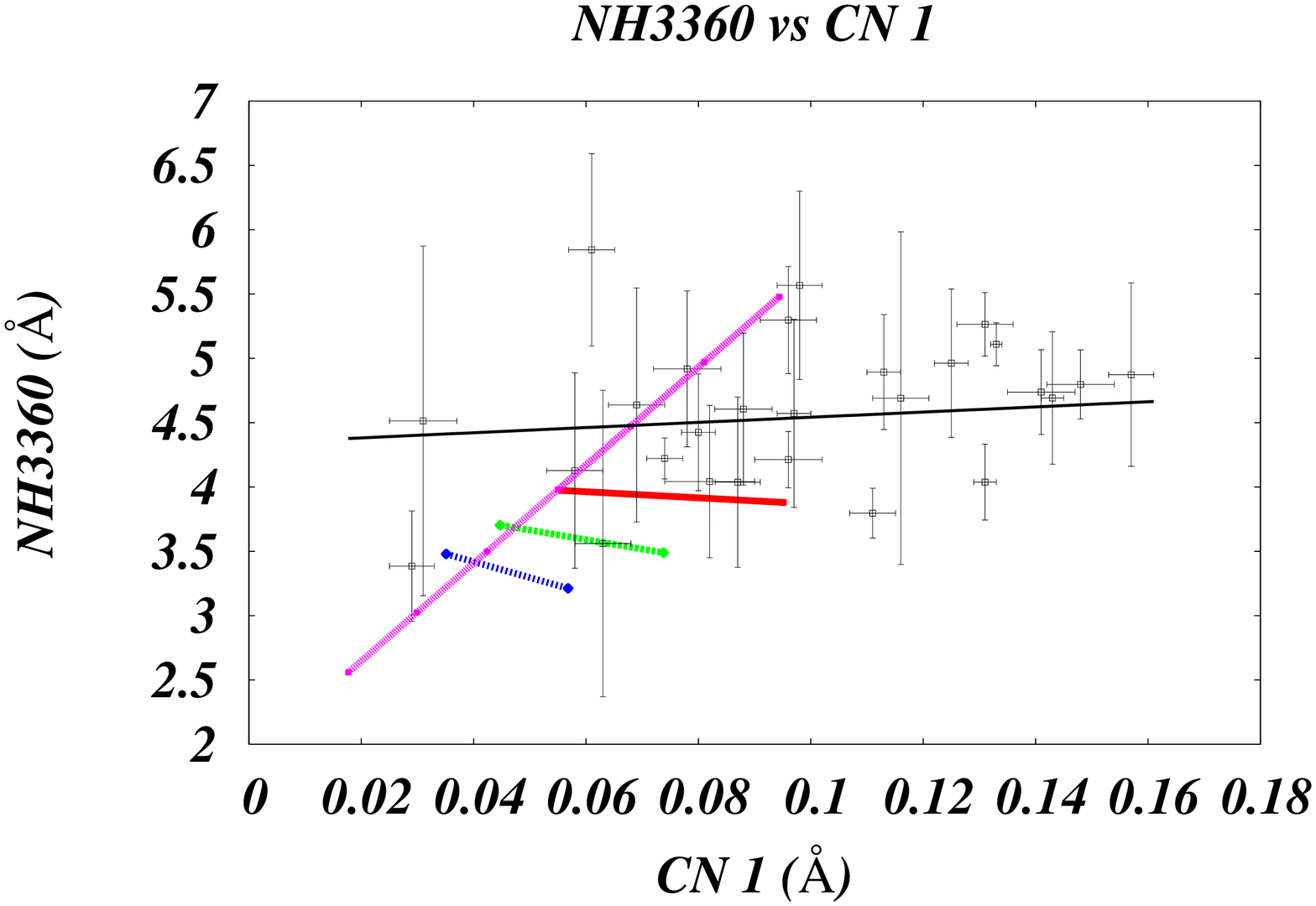}
\caption{The NH3360 and CN$_1$ galaxy indices are plotted. The short lines represent 12 Gyr galaxy models of metallicity [Fe/H] = 0.0 connecting to a metallicity of [Fe/H] = 0.25. The red line represents the models with no subpopulation. The green line represents the same pair of models, but with 5 percent of the galaxy mass in a 12 Gyr, metal-poor ([Fe/H] = $-$1.5) subpopulation. The blue represents the same except that the subpopulation is now 10 percent of the total mass. The pink line represents the same 12 Gyr model of metallicity [Fe/H]= 0.0, but with increasing N abundance from left to right. The left most end is at $- 0.3$ dex, while the right most point is at $+ 0.3$ dex The horizontal black line is a fit to the data. Data as in Fig. 3.
\label{fig4} }
\end{figure}

\begin{table}[H]
\begin{center}
\begin{tabular}{ l c c c}
\multicolumn{4}{c}{Table 3} \\
\hline
\hline
Index &Line Fits &Intercept Error&Slope Error\\
\hline
NH3360  &  $y = 7.85 - 0.59x$ & 2.54&0.46\\
NH3375  &  $y = 6.90 - 0.31x$ & 3.07&0.57\\
Mg3334  &  $y = 0.89 - 0.01x$ & 2.38&0.44 \\
CN 1        &  $y = -0.68 + 0.15x$ &0.17&0.03\\
Mg $b$    &  $y = -4.34 + 1.69x$ &2.19&0.39\\
Mg3334 vs Mg $b$ & $y = 0.64 + 0.05x$ &1.56&0.35\\
H$\beta$ &  $y= 5.94 - 0.82x$ & 6.03&0.68\\
NH3360 vs CN 1 & $y=4.34 + 2.01$ &0.42 & 3.61 \\
\hline
\end{tabular}
\end{center}
\caption{This shows the line fits for the measured indices and the fitting parameters' errors.}
\label{table3}
\end{table}

\begin{table}[H]
\begin{center}
\begin{tabular}{ l r c }
\multicolumn{3}{c}{Table 4} \\
\hline
\hline
Index &Slope&Slope Error\\
\hline

NH3360 Fit &  -0.59  &0.46\\
No Subpopulation& -0.07& \\
$5\%$   & -0.22&\\
$10\%$ & -0.35&\\
NH3375 Fit &  -0.31 &0.57\\
No Subpopulation& 0.38& \\
$5\%$   & 0.15&\\
$10\%$ & -0.04&\\
Mg3334 Fit& -0.01 &0.44 \\
No Subpopulation& 0.29& \\
$5\%$   & 0.22&\\
$10\%$ & 0.16&\\
CN 1      Fit &  0.15  &0.03\\
No Subpopulation& 0.03& \\
$5\%$   & 0.03&\\
$10\%$ & 0.03&\\
Mg $b$  Fit  &  1.69 &0.39\\
No Subpopulation& 0.51& \\
$5\%$   & 0.50&\\
$10\%$ & 0.49&\\
Mg3334 vs. Mg $b$ Fit&  0.05 &0.35\\
No Subpopulation& 0.57& \\
$5\%$   & 0.43&\\
$10\%$ &0.32&\\
H$\beta$ Fit& -0.82 &0.68\\
No Subpopulation& -0.19& \\
$5\%$   & -0.21&\\
$10\%$ & -0.22&\\
NH3360 vs CN 1 fit &   2.01& 3.61   \\
No Subpopulation &-2.37&\\
$5\%$   & -7.36  &\\
$10\%$  & -12.34 &\\
\hline
\end{tabular}
\end{center}
\caption{This shows the line fits' slopes and errors for the measured indices. Also shown are the slopes for the 5 percent and 10 percent by mass metal-poor subpopulation models.}
\label{table4}
\end{table}

\begin{figure}[H]
\includegraphics[width=6.5in]{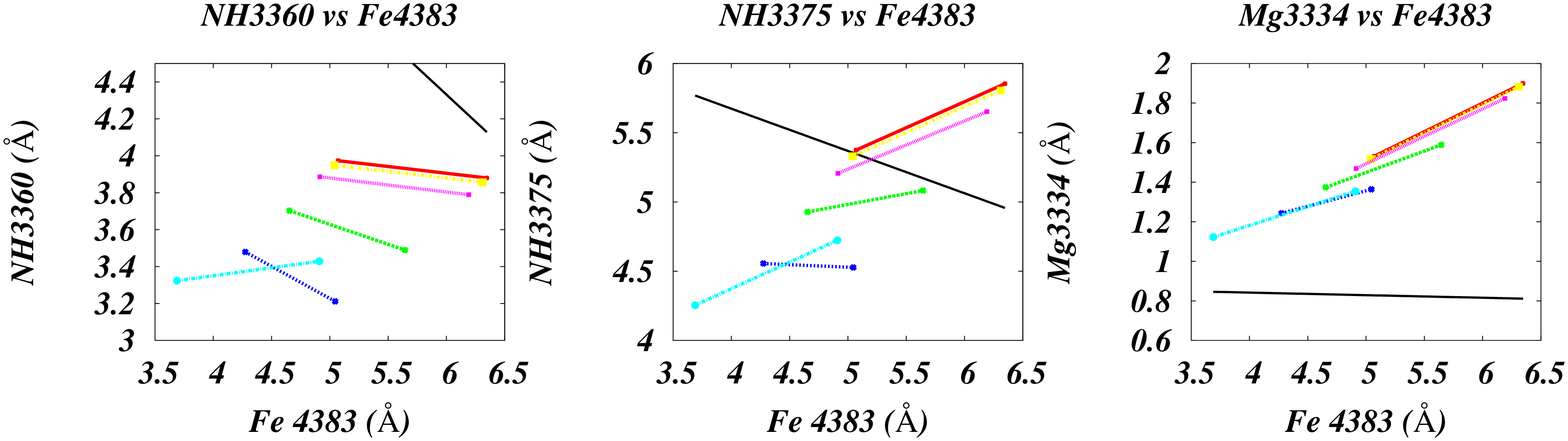}
\includegraphics[width=6.5in]{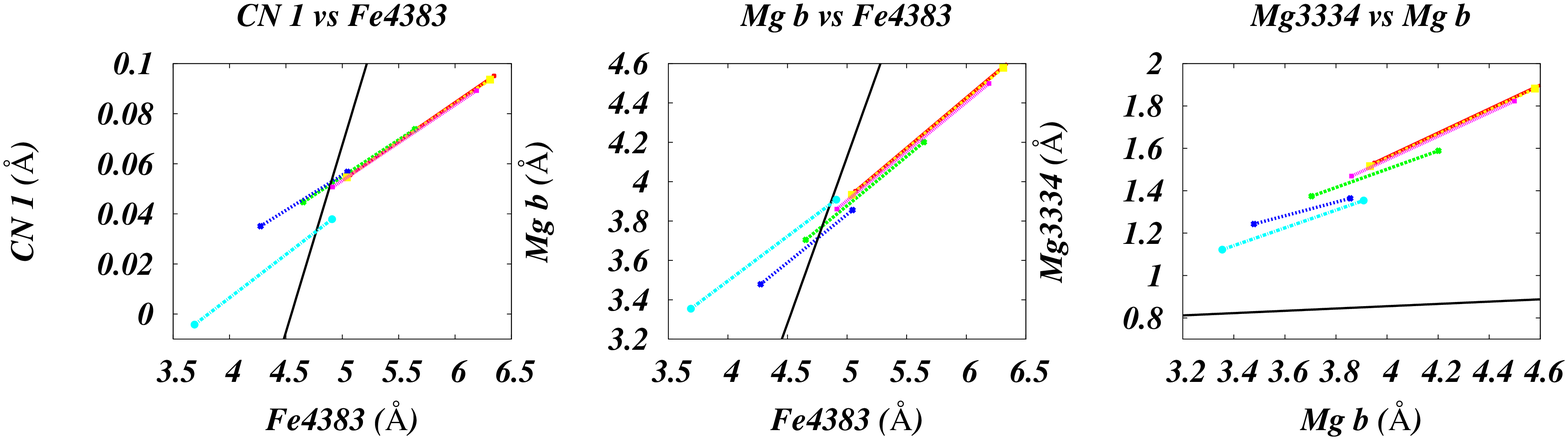}
\includegraphics[width=2.4in]{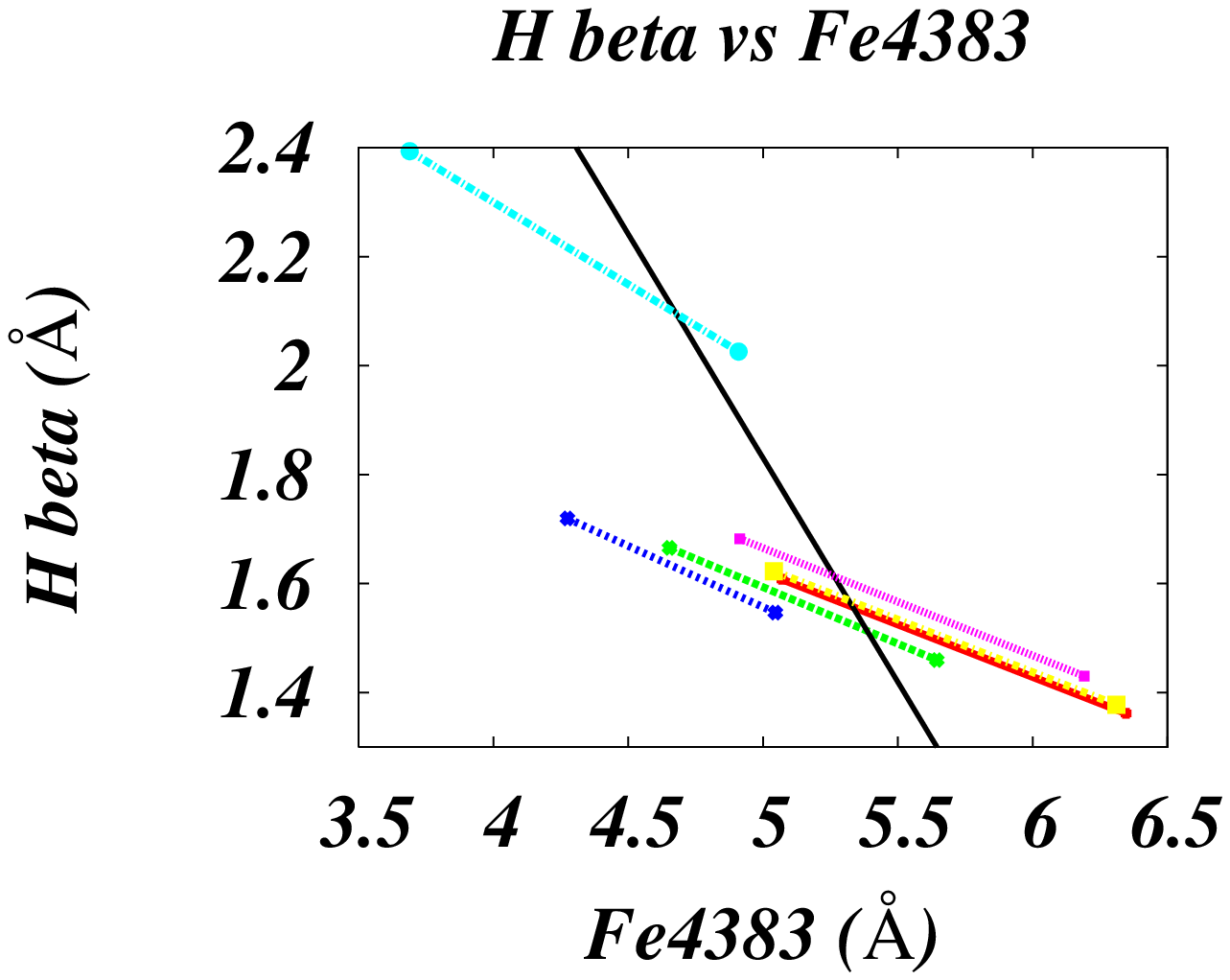}
\caption{In the first, second and third panels, NH3360, NH3375 and Mg3334 indices vs Fe4383 are plotted, respectively. In the fourth and fifth panels, CN$_1$ and Mg $b$ indices vs Fe4383 are plotted respectively. In the sixth panel Mg3334 vs Mg $b$ is plotted and in the seventh panel H$\beta$ is plotted. In each panel along with the index are plotted 7 lines. The lines in red, green and blue represent the same models found in Fig 4. The lines in yellow, pink and light blue represent the respective indices measured from a 12 Gyr galaxy model of metallicity [Fe/H] = 0.0 to a metallicity of [Fe/H] = 0.25 with a 5 percent by mass subpopulation of 8 (yellow), 4 (pink) and 1 (light blue) Gyrs. The black lines are fits to the data.
\label{fig6} }
\end{figure}

To compare the relative effects of a solar metallicity population of varying age on the observed trends. The NH3360, NH3375, Mg3334, CN$_1$, Mg $b$, H$\beta$ indices vs Fe4383 and Mg3334 vs Mg $b$ are plotted (Fig.\ref{fig6}). In each of these plots there are 6 models. The red, green and blue are the models form Figure 4, those consisting of a 12 Gyr model ranging for solar metallicity to [Fe/H] = 0.25 with no subpopulation, a $5\%$ by mass metal-poor 12 Gyr subpopulation and a $10\%$ by mass metal-poor 12 Gyr subpopulation respectively. The remaining three models consist of the 12 Gyr model ranging form [Fe/H] = 0.0 to [Fe/H] = 0.25. The yellow line represents these models with a $5\%$ by mass subpopulation of 8 Gyr. The pink line represents these models with a $5\%$ by mass subpopulation of 4 Gyr and the light blue line represents these models with a $5\%$ by mass 1 Gyr subpopulation. 

These young populations were chosen to investigate any observable effect due the presence of a blue population such as young and intermediate age populations and also blue stragglers whose integrated light mimics that of an intermediate age population (of turnoff mass about twice that of the old population).

What can be seen in Figure 6 and Table 5 is that the addition of a young subpopulation has little to no effect on the index vs. index trends, with the exception of the NH3360 index, in which case only the youngest subpopulation (1 Gyr) has an appreciable effect. The effect of this subpopulation is counter to that of the metal-poor subpopulation and does nothing to help explain the observed trend.

\begin{table}[H]
\begin{center}
\begin{tabular}{ l r c }
\multicolumn{3}{c}{Table 5} \\
\hline
\hline
Index &Slope&Slope Error\\
\hline

NH3360 Fit &  -0.59  &0.46\\
No Subpopulation& -0.07& \\
$5 \%$ 8 Gyr   & -0.07&\\
$5 \%$ 4 Gyr & -0.07&\\
$5 \%$ 1 Gyr & 0.09&\\
NH3375 Fit &  -0.31 &0.57\\
No Subpopulation& 0.38& \\
$5 \%$ 8 Gyr  & 0.37&\\
$5 \%$ 4 Gyr &  0.35&\\
$5 \%$ 1 Gyr &  0.39&\\
Mg3334 Fit& -0.01 &0.44 \\
No Subpopulation& 0.29& \\
$5 \%$ 8 Gyr   & 0.29&\\
$5 \%$ 4 Gyr & 0.28&\\
$5 \%$ 1 Gyr  & 0.19&\\
CN 1      Fit &  0.15  &0.03\\
No Subpopulation& 0.03& \\
$5 \%$ 8 Gyr   & 0.03&\\
$5 \%$ 4 Gyr  & 0.03&\\
$5 \%$ 1 Gyr  & 0.03&\\
Mg $b$  Fit  &  1.69 &0.39\\
No Subpopulation& 0.51& \\
$5 \%$ 8 Gyr   & 0.51&\\
$5 \%$ 4 Gyr & 0.50&\\
$5 \%$ 1 Gyr & 0.45&\\
Mg3334 vs. Mg $b$ Fit&  0.05 &0.35\\
No Subpopulation& 0.57& \\
$5 \%$ 8 Gyr   & 0.57&\\
$5 \%$ 4 Gyr & 0.55&\\
$5 \%$ 1 Gyr & 0.42&\\
H$\beta$ Fit& -0.82 &0.68\\
No Subpopulation& -0.19& \\
$5 \%$ 8 Gyr   & -0.19&\\
$5 \%$ 4 Gyr   & -0.20&\\
$5 \%$ 1 Gyr   & -0.30&\\
\hline
\end{tabular}
\end{center}
\caption{This shows the line fits' slopes and errors for the measured indices. Also shown are the model slopes for 5 percent by mass 8, 4 and 1 Gyrs subpopulations.}
\label{table5}
\end{table}

\section{Discussion, Summary, and Conclusion}

With the use of simple stellar population models we have shown that the NH3360 index, while being C- and O-insensitive, suffers from contamination from other elements, most notably Mg (see Table 6). This dependence of the NH3360 index on Mg can be effectively removed by a simple redefinition of the NH3360 index blue continuum passband to form a new index, index NH3375. This new index is not only insensitive to C and O but also Mg (Table 6) and even though contaminations from other elements exist (mostly Ti), NH3375 is a cleaner measure of N than NH3360.

The predicted behavior of NH3375, as seen in Fig \ref{Fig2}, is an increased sensitivity to N abundance (Table 6). This prediction is not evident in the data (see Fig. \ref{Fig4}). What is evident in the data is that the index trends of NH3375 and NH3360 tend to decrease and the Mg3334 index is nearly flat. This behavior is in contrast to the index trends seen in CN$_1$ and Mg $b$. The premise being violated is that if these indices are measuring N and Mg, respectively, then these indices should show similar behaviors.

There appears to be a difference between the underlying stellar populations that produce the near-UV spectra and the visible spectra at least as regards these near-UV indices. This is plausibly due to a fraction of metal-poor population of the order of 5-10 \% by mass \citep{wor96}. The addition of a metal-poor subpopulation to our models {\em does} account for the discrepancies observed in the index trends (Fig. \ref{Fig3}), most notably those of Mg3334 and Mg $b$. The low metal-poor fractions tested here are of order those inferred by \cite{wor96} for the amount of mass locked into low metallicity stars in present day galaxies.

The presence of a young blue subpopulation has very little effect on the observed index vs index trends. The lone exception is that of NH3360 vs Fe4383 where the effect of a 1 Gyr subpopulation effects the model trends counter to what is needed to explain the observed trends (FIg. 6 and Table 5).

The Mg3334 index turns out to be a useful tool as it is a fairly clean index with very little contamination (Table 6), which allows for a direct comparison to Mg $b$ (Fig \ref{Fig4}). The remarkable flat behavior of Mg3334 compared with Mg $b$ can therefore not be attributed to Mg abundance, but therefore must be due to some other effect, and the metal-poor subpopulations hypothesis explains the difference nicely. Indeed, the pair of the indices taken together can be used to characterize the amount and abundance spread in the underlying metal poor fraction of stellar mass.

With the inherently composite nature of the near-UV spectra \citep{Burstein88}, with metal-poor light mingling with metal-rich light, deriving a N abundance from NH3375 or NH3360 looks more difficult than heretofore suspected. It is outside the scope of this paper, but by using Mg3334 it may be possible to uncover N abundance by comparing derived Mg abundances from Mg3334 and Mg $b$ and then calibrating the underlying metal-poor population until the Mg3334 abundance agrees with that of Mg $b$. This metal-poor calibration could then be applied to NH3375, which in turn could be used to derive a N abundance for comparison with N abundances derived from C, N and O  indices. Hopefully, these measurements will be in agreement with the N enhancements in large elliptical galaxies deduced by other authors \citep{Graves07,Kelson06,wor98a}. 

The existence of a small scatter Mg - $\sigma$ relation among large elliptical galaxies implies increasingly effective Mg enrichment from Type II supernovae with increasing galaxy size. This might be because the initial mass function is stronger in large ellipticals (causing more Type II supernovae and light element enrichment), or the Type Ia delayed timescale is a cause, or because stellar binarism depends on environment \citep{wor98b}. The similarity of the N - $\sigma$ and Na - $\sigma$ relations to that of the Mg - $\sigma$ relation imply that the same mechanism (that of Type II supernovae) is responsible for the observed trends and that contributions from manufacturing on the asymptotic giant branch do not contribute significantly. If N was mostly made in CNO cycle main sequence stars and later ejected, the ejection timescale would be more similar to Type Ia supernovae and N could be expected to track the Fe peak better than light elements. The C - $\sigma$ relation, on the other hand, seems to be intermediate between Mg, N, Na and Fe-peak elements (from Type Ia supernovae; \cite{trag}). This opens a suggestive door to the possibility that the C abundance may come from contributions form both supernovae and the asymptotic giant branch \citep{hen99}.  

\acknowledgements
Major funding for this work was provided by National Science Foundation grants 0307487 and 0346347.

\begin{table}[H]
\begin{center}
\begin{tabular}{ l c c c c c c c c c c c c c}
\multicolumn{14}{c}{Table 6} \\
\hline
\hline
              &  $I_0$ & $\sigma$ & & & & & & & & & & & \\
Index   &   (\AA\/) &     (\AA\/) &     C &       N &      O &      Na &     Mg &    Al &     Si &      S &      K &      Ca &     Sc  \\
\hline
NH3360     &    3.221 &    0.285 &  -0.05 &   5.18 &  -0.52 &  -0.03 &  -3.43 &  -0.04 &  -0.30 &  -0.05 &   0.00 &  -0.20 &   0.13 \\
NH3375     &    4.688 &    0.322 &  -0.20 &   4.51 &  -0.87 &  -0.03 &  -0.71 &  -0.04 &  -0.04 &  -0.05 &   0.00 &  -0.50 &  -0.02 \\
Mg3334      &    1.542 &    0.105 &  -0.43 &  -0.22 &  -1.14 &  -0.04 &   5.91 &  -0.05 &   0.14 &  -0.04 &   0.00 &  -0.35 &  -0.24 \\
Fe4383       &    7.526 &    0.222 &   0.23 &  -0.38 &  -3.09 &  -0.10 &  -1.18 &  -0.35 &  -0.80 &  -0.18 &  -0.02 &  -1.11 &   0.13 \\
H$\beta$    &      1.153 &    0.135 &  -0.04 &   0.13 &   1.55 &   0.24 &  -1.88 &   0.09 &  -0.68 &   0.10 &   0.01 &   0.03 &  -0.07 \\
Mg $b$       &      4.882 &    0.149 &  -1.10 &  -0.33 &  -2.03 &  -0.35 &  10.00 &  -0.21 &  -0.15 &  -0.14 &  -0.01 &   0.02 &  -0.01 \\
\hline
\end{tabular}
\end{center}
\caption{This shows the response of the NH3360, NH3375,  Mg3334 and Fe4383 index definitions to various elements. The first column is the name of the index, the second column gives the index measurements in angstroms of equivalent width, and the third column gives the error associated with S/N = 100 at 5000 \AA\/. The remainder of the columns are the changes (enhanced minus unenhanced) in the index brought on by an element enhancement of 0.3 dex ( or 0.15 dex for C) in units of the error of the third column.}
\label{table7}
\end{table}

\begin{table}[H]
\begin{center}
\begin{tabular}{ l c c c c c c c c c c c c c}
\multicolumn{14}{c}{Table 6 Continued} \\
\hline
\hline
              & & & & & & & & & & & & &  \\
Index   &      Ti &     V &     Cr  &     Mn &     Fe &     Co &     Ni &     Cu &     Zn &    Sr &     Ba &     Eu &    upX2 \\
\hline
NH3360           &   0.04 &   0.37 &  -0.18 &  -0.12 &  -1.70 &  -0.41 &   1.36 &  -0.04 &  -0.37 &   0.01 &   0.00 &   0.00 &  -3.82 \\
NH3375           &  -2.09 &   0.25 &  -0.08 &  -0.25 &  -0.78 &  -0.27 &   1.82 &   0.00 &  -0.57 &   0.00 &   0.00 &   0.01 &  -3.97 \\
Mg3334         &  -0.26 &  -0.52 &   0.18 &  -0.65 &  -0.29 &   0.03 &  -0.92 &   0.06 &  -0.26 &  -0.01 &   0.00 &   0.00 &   3.10 \\
Fe4383          &   0.40 &   0.76 &   0.17 &  -0.57 &   6.97 &  -0.01 &   0.15 &   0.00 &   0.00 &   0.00 &   0.02 &   0.00 &  -6.73 \\
H$\beta$      &   0.33 &  -0.06 &  -0.45 &  -0.10 &  -0.62 &  -0.13 &   0.87 &   0.00 &   0.00 &  -0.02 &   0.00 &   0.00 &   0.40 \\
Mg $b$         &   0.32 &  -0.06 &  -2.20 &  -0.18 &  -1.55 &  -0.07 &  -0.02 &  -0.07 &   0.00 &   0.00 &   0.00 &   0.00 &   6.25 \\
\hline
\end{tabular}
\end{center}
\end{table}

\end{document}